\documentclass[english,usenatbib]{mn2e}
\usepackage[]{fontenc}
\usepackage[latin1]{inputenc}
\usepackage{amsmath}
\usepackage{graphicx}
\usepackage{amssymb}
\usepackage[authoryear]{natbib}

\newcommand{\msun}{\mbox{${\rm M}_\odot$}}

\newcommand{\eq}[1]{\mbox{Eq. #1}}
\newcommand{\fig}[1]{\mbox{Fig. #1}}
\newcommand{\tbl}[1]{\mbox{Tab. #1}}
\newcommand{\sect}[1]{\mbox{Sect. #1}}

\newcommand{\dd}{{\rm d}}

\title{Mixing in massive stellar mergers}

\author[E.Gaburov, J. Lombardi and S. Portegies Zwart]
       {E.\ Gaburov$^{1,2}$, J.\ C.\ Lombardi, Jr.$^{3}$, S.\ Portegies Zwart$^{1,2}$\\
$^1$ Sterrenkundig Instituut ``Anton Pannekoek'', University of Amsterdam, the Netherlands \\ 
$^2$ Section Computational Science, University of Amsterdam, the Netherlands \\ 
$^3$ Department of Physics, Allegheny College, 520 North Main Street, Meadville, PA 16335, USA}

\begin{document}

\maketitle

\begin{abstract}
  The early evolution of dense star clusters is possibly dominated by
  close interactions between stars, and physical collisions between
  stars may occur quite frequently. Simulating a stellar collision
  event can be an intensive numerical task, as detailed calculations
  of this process require hydrodynamic simulations in three
  dimensions.  We present a computationally inexpensive method in
  which we approximate the merger process, including shock heating,
  hydrodynamic mixing and mass loss, with a simple algorithm 
  based on conservation laws and a basic qualitative understanding of
  the hydrodynamics of stellar mergers. The algorithm relies on
  Archimedes' principle to dictate the distribution of the fluid
  in the stable equilibrium situation.  We calibrate and apply the method
  to mergers of massive stars, as these are expected to occur in young
  and dense star clusters. We find that without the effects of
  microscopic mixing, the temperature and chemical composition
  profiles in a collision product can become double-valued functions of
  enclosed mass. Such an unphysical situation is mended by simulating
  microscopic mixing as a post-collision effect. In this way we find
  that head-on collisions between stars of the same spectral type result in
  substantial mixing, while mergers between stars of different spectral type,
  such as type B and O stars ($\sim$10 and $\sim$40\msun
  respectively), are subject to relatively little hydrodynamic mixing.
\end{abstract}

\section{Introduction}\label{sect:Introduction}

The cores of star clusters can become so dense that close
encounters and even collisions between stars become quite common
\citep{1999A&A...348..117P}. This effect is particularly important in
young star clusters where a broad range of stellar masses dominates
the dynamical evolution of the cluster core. In a star cluster with an
initial half-mass relaxation time smaller than a few tens of million
years, core collapse can occur well before the most massive stars
leave the main-sequence and explode in a supernova. In such star
clusters, these massive stars will dominate the core dynamics. This is
most noticeable in both the arrest of core collapse by the formation
of tight massive binaries and the occurrence of physical collisions
between stars which are mediated by the binaries
\citep{2007arXiv0707.0406G}.

Simulating stellar collisions in globular clusters has attracted
considerable attention over the last decade. These calculations were
mainly initiated to explain the presence and characteristics of blue
stragglers
\citep{2002MNRAS.332...49S,2002ApJ...568..939L,1996ApJ...468..797L}.
In many cases, collisions occur infrequently enough that the already
computationally expensive dynamical modelling of a cluster would
hardly be slowed down by a simple treatment of collisions. However, it
was recently recognised that the number of collisions in young star
clusters can become quite large, and this may lead to formation of a
very massive object \citep{2004Natur.428..724P,
  2006MNRAS.368..141F}. Furthermore, the subsequent evolution of such
objects, and of collision products in general, can depend on their
internal structure.

To determine the structure of a collision product accurately, one can
resort to high-resolution numerical simulations of stellar
mergers. Such simulations are generally carried out by means of the
smoothed particle hydrodynamics (SPH) method
\citep{0034-4885-68-8-R01}. This however becomes computationally
expensive, especially if one is interested in multiple collisions
which may occur in the course of $N$-body simulations. On the other
hand, components of multiple systems, such as binaries and triples,
may experience a collision \citep{2007arXiv0707.0406G}, and careful
treatment of such collisions might play an important role for further
evolution of these multiple systems. Therefore, faster methods are
needed to compute the outcome of a stellar merger event, and these
methods can eventually be included in $N$-body simulations.

\cite{1996ApJ...468..797L} constructed a method which has been
successfully used for collisions between low-mass main-sequence stars,
and in this letter we extend this method to collisions between massive
stars. Our method has a general applicability, in the sense that it is
suitable for mergers of all stellar types, including compact objects.

 \section{Methods}\label{sect:Methods}

\subsection{Guiding Principle}

The guiding idea of our method is Archimedes' principle.  To determine
how fluid will redistribute itself when brought out of equilibrium,
perhaps by a stellar merger, we need to consider densities: a fluid
element with a greater density than its environment will accelerate
downward, while one with a smaller density will be buoyed upward.  Any
such element can ultimately settle into hydrostatic equilibrium once
its density equals that of its environment.  The density of a fluid
element will not in general remain constant during the collision.
Instead, it will be continuously adjusted in such a way that pressure
equilibrium with the environment is achieved.  In other words, parcels
of gas expand or contract as necessary to equilibrate their pressure
with the environment.

We begin by considering a distribution of gas, in our case stars, and
specify distribution functions for the pressure $P$, density $\rho$,
and abundance of chemical species $X_i$ as a function of the enclosed
mass.  For a given equation of state, the expression for the entropy
can be determined.  It is often mathematically simpler, although not
formally necessary, to define an entropic variable $A$ which is
related to the specific entropy and conserved by fluid elements in
adiabatic processes; we call this entropy variable $A$ the buoyancy.

The buoyancy can be calculated from the equation of state,
$P=P(A,\rho,X_i)$.  In the case of an ideal gas $A= P/\rho^\Gamma$,
where $\Gamma$ is the adiabatic index. By construction, the buoyancy
$A$ depends directly upon entropy and composition, and it remains
constant for each fluid element in the absence of heating and
mixing. Therefore, it is an important starting point for understanding
the hydrodynamics of collisions.

This idea has been successfully used to describe the underlying
hydrodynamics of collisions among low mass main sequence stars
\citep{1996ApJ...468..797L,2002ApJ...568..939L,2003MNRAS.345..762L},
which are well described by a monatomic ideal gas equation of state.
In this case $\rho=(P/A)^{3/5}$, such that at a given pressure, a
smaller $A$ yields a greater value of $\rho$.  Consequently, fluid
with smaller values of $A$ sink to the bottom of a potential well, and
the $A$ profile of a star in stable hydrostatic equilibrium increases
radially outwards. Indeed, it is straightforward to show that the
condition ${\rm d}A/{\rm d}m>0$ is equivalent to the usual Ledoux
criterion for convective stability of a nonrotating star
\citep{1996ApJ...468..797L}.

% For a detailed discussion of the stability conditions within
% rotating stars, see \S 7.3 of \citet{1978trs..book.....T} or
% \citet{2000stro.book.....T}.

%% During a collision, fluid from the parent stars is shock heated. This
%% can be imitated by increasing the buoyancy of the fluid in order to
%% yield a final total energy which is equal to the initial total energy
%% of the system.

%% There are, however, astrophysically relevant scenarios in which the
%% relative impact speed of two colliding stars does not greatly
%% exceed the speed of sound inside them. Consequently, the resulting
%% shocks have Mach numbers of order unity and shock heating is
%% relatively weak. In such cases, to a reasonable approximation,
%% fluid elements each maintain a constant $A$ throughout a collision.
%% Therefore, as described in more detail below, the distribution of
%% fluid in a collisional product can be approximately determined
%% simply by sorting the fluid from both parent stars in order of the
%% adiabatically mapped density.  The model can be further refined by
%% including the effects of shock heating in a way constrained by
%% energy conservation: the fluid must be heated during the collision
%% by the amount necessary to yield a final total energy equal to the
%% initial total energy of the system.

\subsection{Sorting Algorithm}

The fluid in high mass main-sequence stars is well described by a
mixture of monatomic ideal gas and radiation in thermal equilibrium.
In such cases, the total pressure is
\begin{equation}
  P = \frac{\rho k T}{\mu} + \frac{a T^4}{3}, \label{eq:pressure}
\end{equation}
where $k$ is the Boltzmann constant, $T$ is temperature, $\mu$ is the
mean molecular mass, and $a$ is the radiation constant.  The specific
entropy is \citep{1983psen.book.....C}
\begin{equation}
  s - s_0 = {3\over 2}{k\over \mu} \left[
  \log\left({kT\over \rho^{2/3}\mu}\right)+{8\over 3}\frac{1-\beta}{\beta}\right] \equiv
  {3\over 2 }{k \over \mu} \log A, \label{eq:entropy}
\end{equation}
where the quantity $s_0$ depends only on composition and $\beta$ is
the ratio of gas to total pressure.
% A fluid element of given composition therefore experiences a change
%in specific entropy from a final state $f$ to an initial state $i$
%of \begin{eqnarray} \Delta s & = & {k\over \mu} \left\{\log\left[
%\left({T_f\over T_i}\right)^{3/2} {\rho_i\over \rho_f}\right]
%+4\left(\frac{1-\beta_f}{\beta_f} - \frac{1-\beta_i}{\beta_i}\right)
%\right\} \nonumber\\ & = &{3\over 2}{k \over \mu} \log {A_f\over
%A_i}, \end{eqnarray} where
To achieve the equality in \eq{\ref{eq:entropy}}, we define the
buoyancy $A$ through the following relation
\begin{equation}
  P = {A\rho^{5/3}\over \beta}e^{-{8\over 3}{1 -\beta \over
      \beta}}. \label{eq:A}
\end{equation}
%
%By solving eq.\ (\ref{A}) for $A$ and substituting into eq.\ (\ref{entropy}) it is straightforward to verify that eq.\ (\ref{clayton}) is recovered.
%
Through manipulations with \eq{\ref{eq:pressure}},
\eq{\ref{eq:entropy}}, \eq{\ref{eq:A}} and definition of $\beta$, we
find
\begin{equation}
  \rho = {3 k^4 \over A^3 \mu^4 a}{1 - \beta \over \beta}
  e^{8{1-\beta \over \beta}}. \label{eq:eos_rho}
\end{equation}
Together, \eq{\ref{eq:A}} and \eq{\ref{eq:eos_rho}} provide the
desired relationships between the pressure, buoyancy, density, and
composition.  Given a prescription for shock heating and an estimate
of the mass loss, it is possible to determine the structure of a
non-rotating product resulting from a head-on collision of two massive
main-sequence stars, which we call parent stars.

To begin with, we consider fluid elements at the centre of each parent
star and determine which one will settle in the centre of the
collision product. For this purpose, we both make an initial estimate
of the central pressure of the collision product and test which fluid
element would have larger density when brought to this central
pressure: a process that can be completed in two steps.  First, as the
composition and post-shock buoyancy of each fluid element can be
considered known, \eq{\ref{eq:A}} gives what would be the final
$\beta$ value of each element if it were to settle at the centre of
the collision product.  Second, these $\beta$ values correspond to
densities through \eq{\ref{eq:eos_rho}}. The fluid element with the
greater density is the one that actually settles in the centre of the
collision product.  Next, we integrate the equations of hydrostatic
equilibrium to determine the local pressure $P$ throughout the
product. This local pressure is used to determine which of subsequent
fluid elements from the parent stars contribute to the structure of
the product. This procedure is repeated until the mass supply of the
parent stars is depleted. As in the case of low-mass main-sequence
stellar collisions, the central pressure is iteratively improved until
the outer boundary condition, namely that the surface pressure
vanishes, is satisfied.
%% +++ v1 +++
%As the initial estimate of the central pressure of the collision
%product, we use the central pressure of the most massive parent star.
%% +++

As a consequence of this merging procedure, one must expect that fluid
of a certain composition that has originated in one parent star can be
located arbitrarily close to fluid of a different composition from the
other parent.  These fluid elements must be in pressure equilibrium
and also, by the condition of hydrostatic equilibrium, have the same
density.  Therefore, in order to have adjacent fluid elements of
different $\mu$, \eq{\ref{eq:A}} and \eq{\ref{eq:eos_rho}} imply that
these elements must have different buoyancies $A$.  We therefore
should not expect that the $A$ and $\mu$ profiles in a nascent
collision product will be single-valued, at least not before mixing
has occurred on the microscopic level.

If the parent stars have essentially the same homogeneous composition
as in the merger of two unevolved stars, then $\mu$ will be uniform
throughout the collision product, and for any given pressure, greater
densities will correspond to smaller values of the buoyancy.  In other
words, the structure of the collision product can then be simply
determined by sorting the fluid in order of increasing $A$: the lowest
$A$ fluid is placed at the centre of the product and is surrounded by
shells with increasingly higher values of $A$. The merging procedure
also reduces to sorting by $A$ in another special case, namely when
radiation pressure is negligible ($\beta \rightarrow 1$), as
\eq{\ref{eq:A}} clearly reduces to the monatomic ideal gas result in
this limit. The merging procedure presented here is therefore a
natural generalization of that presented for low mass stars in
\citet{2002ApJ...568..939L,2003MNRAS.345..762L}.

% In the discussion that
% follows, we can therefore assume that our iterative algorithm has
% converged upon the correct central pressure.  Everything else is just
% like in earlier MMAS version:  We keep working  our way out through
% the star until m=M.  If at that point the pressure P  has dropped to
% zero, then we have a good model.  But if at m=M we find  P>0, then our
% central pressure guess was to large... so we decrease the  guess and
% try again.  If P<0 at m=M, we increase the central pressure  guess and
% try again.

%The lowest $A$ fluid from the parent
%stars is placed at the core of the remnant and is surrounded by shells
%with increasingly larger $A$.

\subsection{Stability Criterion in High Mass Stars}

The unusual profiles that can be achieved in stellar mergers suggest
that we take a careful look at the condition for convective stability:
\begin{equation}
  \left({\dd\rho \over \dd r}\right)_{\rm env} < 
  \left({\dd\rho \over \dd r}\right)_{\rm el}.
  \label{eq:stability_general}
\end{equation}
Here, the subscripts ``env'' and ``el'' refer to the environment and
the fluid element respectively. Since pressure equilibrium between the
element and its immediate environment is established nearly
instantaneously, the condition for convective stability can be written
in the following way:
\begin{equation}
  \left({\dd\log P \over \dd\log\rho}\right)_{\rm env} <
  \left({\dd\log P \over \dd\log\rho}\right)_{\rm el}. \label{eq:stability}
\end{equation}
Using \eq{\ref{eq:A}} and \eq{\ref{eq:eos_rho}}, we evaluate
derivatives on both sides of \eq{\ref{eq:stability}}. We then note
that
\begin{equation}
  \Gamma_{1}=\frac{-3\beta^{2}-24\beta+32}{24-21\beta}= \frac{5}{3} -
  \frac{3\beta^{2}-11\beta+8}{24-21\beta},
\end{equation}
and this allows us to write the stability condition in the following
form
\begin{equation}
  \frac{3\Gamma_{1}-4}{4}
    \left(\frac{\dd\log A}{\dd\log\rho}\right)_{\rm env} <
    \frac{5 - 3\Gamma_{1}}{3} 
    \left(\frac{\dd\log\mu}{\dd\log\rho}\right)_{\rm
    env}.
\end{equation}
As the $\dd\log \rho/\dd\log m$ is always negative throughout the
star, we can rewrite the stability condition in a more convenient
form:
\begin{equation}
  \left(\frac{\dd\log A}{\dd\log m}\right)_{\rm env} > 
  \frac{4}{3}\frac{\frac{5}{3} - \Gamma_{1}} {\Gamma_{1} - \frac{4}{3}} 
  \left(\frac{\dd\log\mu}{\dd\log m}\right)_{\rm env}.\label{eq:stability2}
\end{equation}
In a star in which \eq{\ref{eq:stability2}} is satisfied, a perturbed
element will experience restoring forces that cause it to oscillate
around its equilibrium position.  It can be seen that in the limit of
ideal gas ($\Gamma_{1}=5/3$), or when the composition is uniform, we
recover the usual stability criteria $\dd A/\dd m > 0$.  In the
general case, however, $A$ will typically increase outward, but can
decrease in regions with an inverted composition gradient.

 \section{Validation}\label{sect:Validation}

\subsection{Initial conditions}
In the previous section we have presented a sorting algorithm that
generates a collision product in hydrostatic equilibrium. As the
method deals with non-rotating products, we validate it by carrying
out simulations of head-on collisions between main-sequence stars of
different mass and age. The full set of simulations is presented in
\tbl{\ref{tbl:simulations}}.
\begin{table}
  \begin{center}
    \begin{tabular}{ccccc}
      \hline
      Model & $M_1$ & $M_2$ & Evolutionary State & $N$   \\
      \hline
      T88, H88      & 80    & 8    & TAMS, HAMS       & 880k \\
      T48, H48      & 40    & 8    & TAMS, HAMS       & 480k \\
      T28, H28, Z28 & 20    & 8    & TAMS, HAMS, ZAMS & 280k \\
      T18, H18, Z18 & 10    & 8    & TAMS, HAMS, ZAMS & 180k \\
      \hline
    \end{tabular}
    \caption{The simulations carried out in this work. The model name
      of the simulations are given in the first column. The masses of
      the primary and the secondary stars are shown in the second and
      the third column respectively. The evolutionary state of the
      parent stars is given in the fourth column: TAMS, HAMS, and ZAMS
      stand for turn-off age, half-age, and zero-age main-sequence
      respectively. In the fifth column we show the number of SPH
      particles in the simulations; in all cases, we
      use well over 100k equal mass SPH particles in order to achieve high
      resolution in the dense stellar interiors and 
      to insure convergent results.}
    \label{tbl:simulations}
  \end{center}
\end{table}

Simulations of stellar collisions were carried out by means of a
modified version of the {\tt GADGET2} code
\citep{2005MNRAS.364.1105S}. In particular, we included radiation
pressure in the equation of state and the functionality which allows
generation of relaxed models of parent stars in quasi-hydrostatic
equilibrium \citep{2006ApJ...640..441L, 1995MNRAS.277..705T}.

We have prepared three-dimensional SPH stellar models from
one-dimensional stellar models computed with the {\tt EZ} stellar
evolution code \citep{2004PASP..116..699P}. In particular, we have
used {\tt EZ} to evolve a non-rotating zero age main-sequence (ZAMS)
model with a primordial helium abundance $Y\simeq 0.3$ and metallicity
$Z=0.02$.  The resulting star is composed of spherical mass-shells
which store necessary stellar data, such as local composition,
temperature, and density, as well as enclosed stellar mass and radius. The
stars are evolved until the primary star reaches its half-age of its
main-sequence life-time (HAMS) or until the primary has only 1\%
hydrogen abundance in its core (TAMS).  In all cases the secondary
star is evolved to the same age as the primary with which it collides.

\subsection{Results}

\begin{figure}
  \begin{center}
    \includegraphics[scale=0.45]{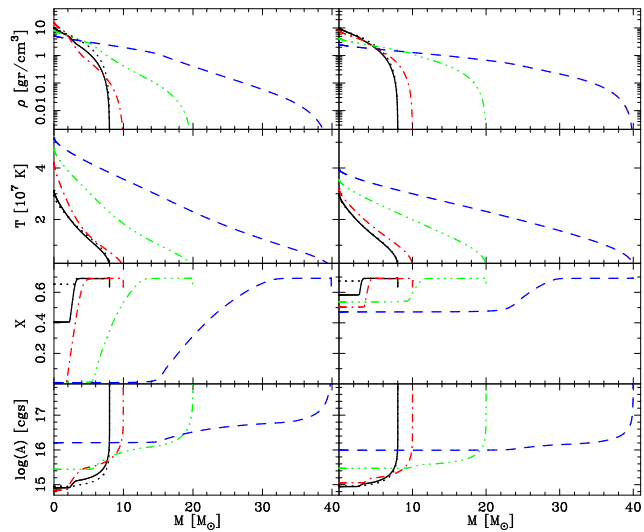}
  \end{center}
  \caption[]{Internal structure of a selection of TAMS (left side) and
    HAMS (right side) parent stars which were used in simulations. The
    upper panels show density profile as a function of enclosed
    mass. The black solid and dotted lines on the left side show
    density profile of the 8\msun\, star from T18 and T48 models
    respectively, while on the right side the 8\msun\, star is taken
    from H18 and H48 models. The red dash-dotted, green
    dash-dot-dot-dot and blue dashed lines show density profiles of
    10\msun\,, 20\msun\, and 40\msun\, stars respectively. The second
    panels from the top show temperature profiles, while the third and
    fourth panels present hydrogen mass fraction and buoyancy profiles
    respectively.}
  \label{fig:parent_stars}
\end{figure}

We have shown in \sect{\ref{sect:Methods}} that the two main
quantities which determine the structure of the collision product is
buoyancy and composition. Since in main-sequence stars the mean
molecular mass is mostly dependent on hydrogen mass fraction, we
present in \fig{\ref{fig:parent_stars}} buoyancy and hydrogen mass
fraction profiles for a selection of the parent stars. In the case of
TAMS stars, the primary consumed nearly all of hydrogen in its
core. Nevertheless, the secondary has consumed less than the half of
its initial hydrogen supply. HAMS parents, on the other hand, are
barely evolved; all of the parent models have burnt less than the half
of the initial amount of hydrogen in their cores.

The buoyancy of the parent stars is shown in the two lowest panels of
\fig{\ref{fig:parent_stars}}. Even though the 20\msun\, and 40\msun\,
stars are at the end of the main-sequence, the buoyancy in their cores
is noticeably higher than that of the secondary star. Therefore, based
on the results of \sect{\ref{sect:Methods}}, it is natural to expect
that the 8\msun\, star will occupy the centre of the merger
product. In the case of the 10\msun\, primary, however, one may expect
mixing, since the buoyancy of the primary and the secondary stars
differ by less than a factor of two, and this difference can be
modified by shock heating.

In this work, we model shock heating by increasing the buoyancy of
each fluid element. For this we need to assume a particular
distribution of shock heating. Motivated by the results of
\cite{2002ApJ...568..939L}, we adopt the following simple
distribution:
\begin{equation}
  \log(A_{\rm f}/A_{\rm i}) = a + b\cdot
  \log(P_{\rm \,i}/P_{\rm c, i}). \label{eq:shock_heating}
\end{equation}
Here, $A_{\rm i}$ and $A_{\rm f}$ are the initial and the final
buoyancy of a fluid element, respectively, $P_{\rm i}$ is the initial
pressure, and $P_{\rm c,i}$ is the central
pressure of a parent stars from which the fluid has come. We
compute parameters $a$ and $b$ by fitting \eq{\ref{eq:shock_heating}}
to the simulation data.

One of the advantages of using \eq{\ref{eq:shock_heating}} can be seen
by looking at \eq{\ref{eq:A}}. As an SPH particle preserves its
composition in the course of a simulation, \eq{\ref{eq:shock_heating}}
describes the change in entropy due to the shock heating
process. This, in turn, potentially expands the range of applicability
of the sorting algorithm presented in \sect{\ref{sect:Methods}} to an
arbitrary equation of state, such as of degenerate matter inside white
dwarfs.

For every simulation from \tbl{\ref{tbl:simulations}}, we obtain $a$
and $b$ values in \eq{\ref{eq:shock_heating}} for each of the parent
stars. We use this equation to compute the change in the buoyancy for
each of the mass shells of the parent stars, and then we apply the
sorting algorithm to generate the structure of the collision product
which we compare with the simulation results.

\begin{figure*}
  \begin{center}
    \includegraphics[scale=0.45]{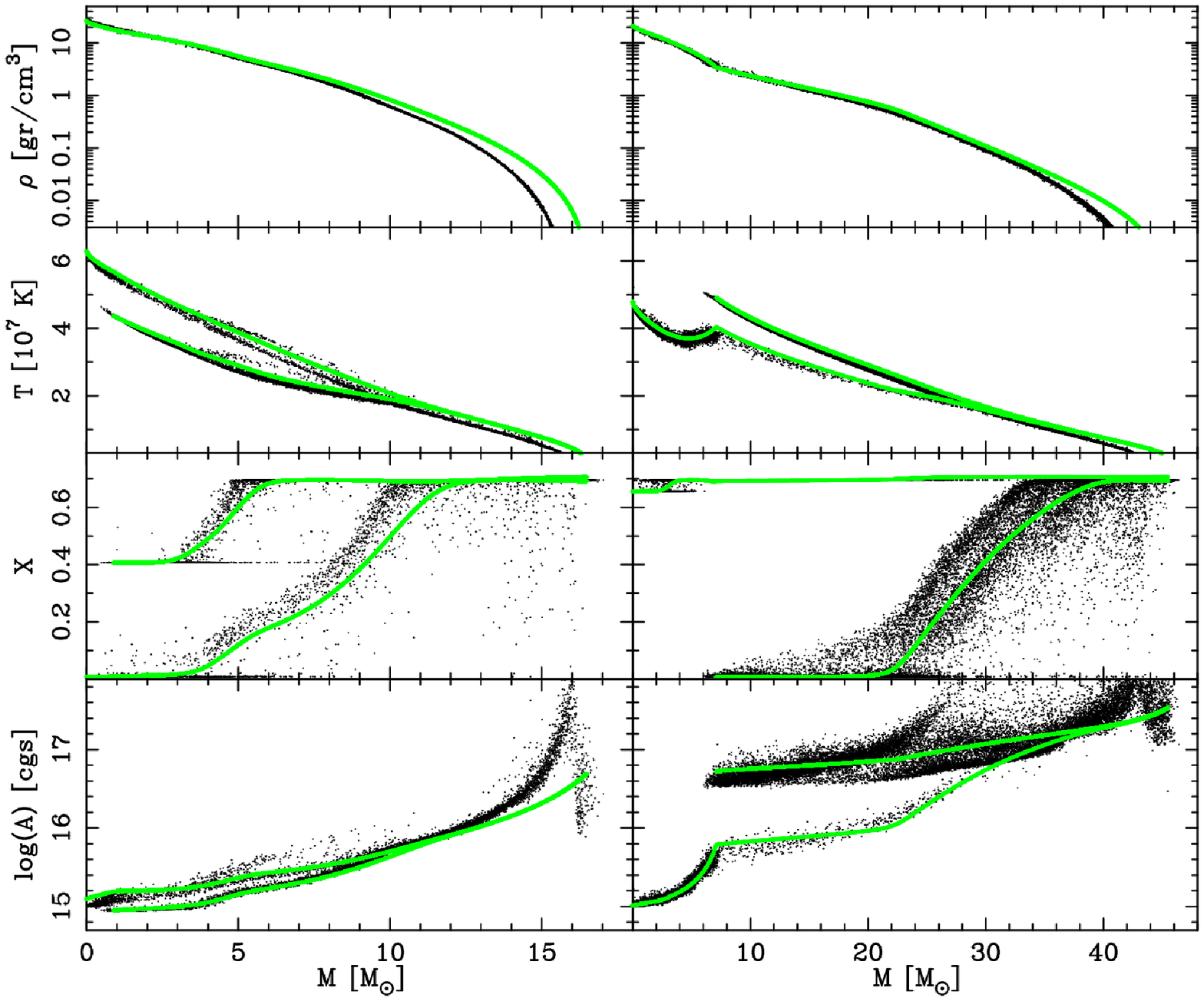}$\,\,\,$
    \includegraphics[scale=0.45]{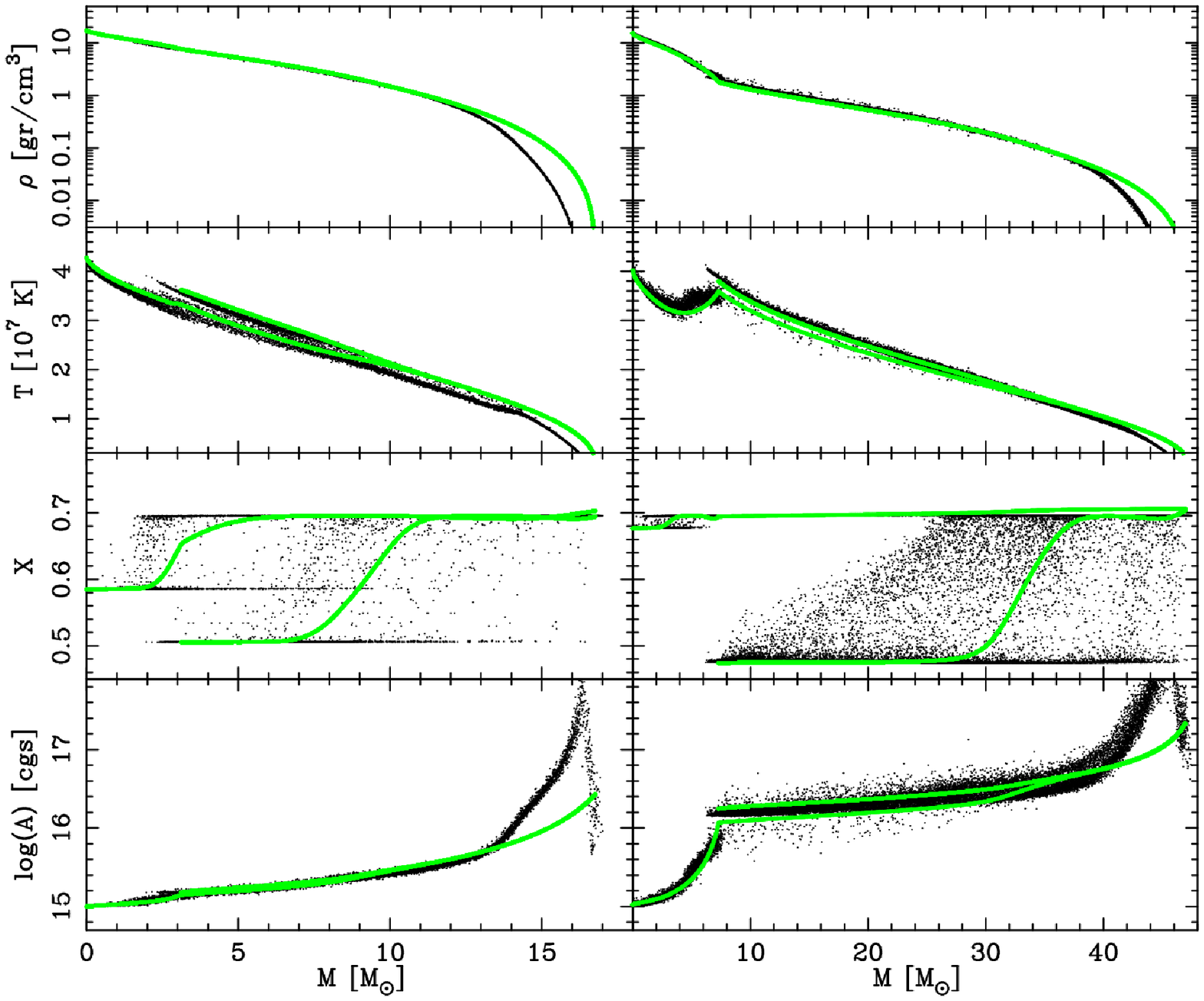}$\,\,\,$
  \end{center}
  \caption[]{The non-mixed structure of the collision products from a
    selection of simulations. The left set of plots displays results
    for TAMS collisions, whereas the right set displays the same data
    for HAMS collisions. The black dots are the results of
    simulations, while the green circles are the results of the
    sorting algorithm. The upper panels show the density profile as a
    function of enclosed mass for T18, T48, H18 and H48 simulation
    models from left to right respectively. In the second panels from
    the top, we show the temperature profile, and in the third and the
    fourth panels, we show the hydrogen mass fraction and buoyancy
    profiles respectively.}
  \label{fig:unmixed}
\end{figure*}

In \fig{\ref{fig:unmixed}} we present the results of the sorting
algorithm and simulations. The sorting algorithm is able to reproduce
two distinct branches in temperature, composition and buoyancy for
TAMS collisions; the density, on the other hand, is single valued as
one would naturally expect. In the case of HAMS collisions, the
branches are less pronounced, as expected, because the parent stars
have roughly the same composition (see
\fig{\ref{fig:parent_stars}}). A noticeable discrepancy in the density
of the outer layers is caused by an inadequate representation of shock
heating in these layers by \eq{\ref{eq:shock_heating}}; indeed, both
the density and buoyancy of our simple model diverge from the SPH
results at about the same enclosed mass.
%%
%% +++ v1
The mass loss in the collisions from \tbl{\ref{tbl:simulations}} never
exceeds 10\%: T18(8.3\%), T28(8.9\%), T48(5.0\%), T88(1.9\%),
H18(6.8\%), H28(4.7\%), H48(2.1\%), H88(0.8\%), Z18(6.4\%) and
Z28(4.5\%). As we expect, mass loss is less in collisions involving
stars of significantly different mass, as less kinetic energy is redirected into
the ejecta. In addition, we observe that mass loss is larger in
collisions with evolved stars, due to their weakly bound envelopes.
%% +++
%%

We note that the very existence of two branches in the results of
Fig.\ \ref{fig:unmixed} means that the buoyancy $A$ {\it does not}
increase outward at all locations.  In the case of model T18, it is
actually the higher buoyancy branch, corresponding to hydrogen
depleted fluid from the TAMS primary, which extends into the core of
the merger product. Such a configuration is stable only because it is
accompanied by a sufficiently steep gradient in composition.  Indeed,
in model H18, the parent stars are too young to yield large gradients
in composition, and consequently it is the lower buoyancy fluid from
the secondary that sinks to the centre of the merger product.  It is
also worth noting that in models Z18 and Z28, in which the ZAMS parent
stars have uniform composition, we find that the buoyancy increases
outwards throughout the collision products, as expected from
\eq{\ref{eq:stability2}}.

As we increase the resolution of parent stars as treated by the
sorting algorithm, adjacent mass shells in different branches become
increasingly closely spaced. The resulting large gradients will
eventually be smoothed by the microscopic mixing between neighbouring
fluid elements. To simulate such mixing, we process the results from
the sorting algorithm into equal-mass bins, such that each bin
accommodates a sufficiently large number of mass shells. We then
average the stellar data, such that the total volume and thermal
energy of each bin, as well as the total mass of each of the chemical
elements, are conserved. From the average values of density, thermal
energy and composition, we compute the rest of the thermodynamic
quantities, such as temperature. We find that this procedure converges
once the high resolution model has roughly $10^5$ mass shells averaged
into 200 bins.  The simulation data is averaged in the same way.

\begin{figure}
  \begin{center}
    \includegraphics[scale=0.44]{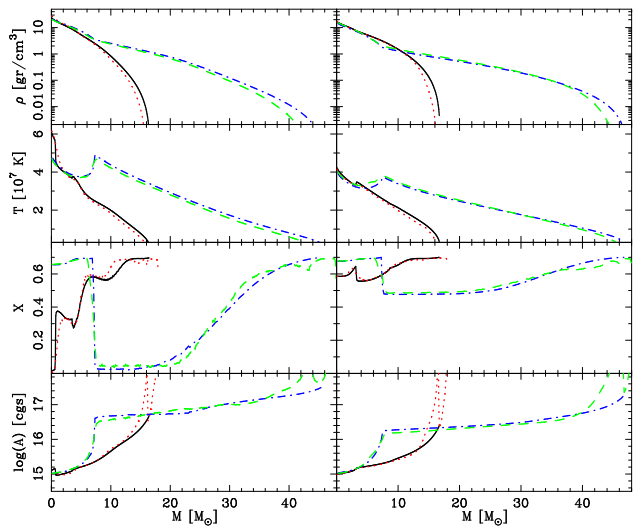}
  \end{center}
  \caption[]{The structure of the mixed collision products from the
    same selection of models as presented in
    \fig{\ref{fig:unmixed}}. Here, the black solid and red dotted
    lines show the results of the sorting algorithm and of the
    simulation respectively for the cases T18 on the left side and H18
    on the right side. The blue dash-dotted and green dashed lines
    show the results of sorting algorithm and of the simulation
    respectively for the cases T48 on the left side and H48 on the
    right side. The upper panels show density profiles as a function of
    enclosed mass, and the second panel from the top shows
    temperature profiles. The third and the fourth panels from the top
    show hydrogen mass fraction and buoyancy profiles respectively.}
  \label{fig:mixed}
\end{figure}

In \fig{\ref{fig:mixed}} we show the results of the mixing applied to
the models shown in \fig{\ref{fig:unmixed}}. Whereas no significant
mixing occurs in models T48 and H48 as the $8\msun$ star settled in
the centre of the primary, the interior of the collision product in
model T18 is mixed. In model H18, however, only the core of the
secondary star settles in the centre of the primary star; the rest of
the fluid is mixed. The resulting buoyancy increases outwards
throughout the product in all models, except in the very centre of
T18. Here, the decrease in buoyancy at $\simeq 0.8\msun\,$ is
accompanied by an increase in mean molecular mass in such a way that
\eq{\ref{eq:stability2}} is satisfied.

\section{Discussion and conclusions}\label{sect:Discussion}
We have presented a new sorting algorithm which generates the
structure of collision products in hydrostatic equilibrium. We have
tested the algorithm by carrying out a set of SPH simulations of
head-on stellar collisions between massive main-sequence stars of
different masses and ages. By calibrating the shock heating with the
simulation data, we have been able to quickly generate the structure
of the collision products consistent with the results of the
simulations.

It takes less than a minute to generate a collision product for any of
the models presented in \tbl{\ref{tbl:simulations}}, whereas an SPH
simulations takes at least a day of CPU time on the same
computer. Such a speed-up will help to make it possible to include
realistic stellar collisions in simulations of star clusters.

We have found that the chemical composition and temperature can be a
multivalued function of the enclosed mass, if microscopic mixing
processes are ignored. This is exhibited by the fact that the
neighbouring mass shells have a discontinuously changing temperature
and chemical composition, whereas the density and pressure remains
continuous. Using the sorting algorithm, we have found that the
situation is unaffected by increasing the number of shells. Instead,
the chemical composition and temperature become double-valued in the
continuous limit. Such an obviously unphysical situation is mended by
microscopic mixing. We have simulated the mixing process by
conservatively averaging multiple mass shells into a single mass
bin. The resulted profiles are both single valued and consistent with
the simulation data.

We have found significant mixing in only the T18 and H18 models. The
reason for this is that buoyancy in the core of the primary and the
secondary stars are similar, whereas for the other models, such as T48
and H48, the buoyancy of the primary stars is substantially larger
than that of the secondary stars. Therefore, in those cases, fluid
from the secondary star occupies the central region of the collision
product without being mixed with the fluid from the primary star. We
therefore conclude, that hydrodynamic mixing in collisions between
main-sequence stars occurs only when the masses and ages of the
parents stars are similar to each other. We find that mixing already
becomes unimportant at the mass ratio $q\simeq 0.4$, which is the mass
ratio of models T28, H28 and Z28. In such cases, the secondary star
simply occupies the centre of the collision product without much
mixing \citep{2006MNRAS.366.1424D, 2007astro.ph..3290S}.

Microscopic mixing may play an important role during the collision
process itself. If the diffusion processes are ignored, the turbulent
motion of the fluid can result in a discontinuous distribution of
chemical composition and temperature, while maintaining continuity in
density and pressure. Such a distribution is susceptible to
microscopic mixing which tends to smooth the discontinuities
\citep{2005ARFM...37..329D}. If the timescale of microscopic mixing is
larger than the time a collision product takes to settle into
quasi-hydrostatic equilibrium, the mixing can be ignored during the
collision event and applied only as a post-collision process. On the
other hand, if the timescale of the microscopic mixing is shorter than
the timescale of a collision event, then one must include the effects
of mixing into the simulation itself.

The sorting algorithm can in principle be applied to any type of
stars; for example, it is possible to apply it to the merger of white
dwarfs. The basic requirement is simply to have the equation of state
which gives the relationship between pressure, density, composition,
and specific entropy (or buoyancy), analogous to Eqs.\ \ref{eq:A} and
\ref{eq:eos_rho}.  While it is also necessary to determine mass loss
and shock heating of the fluid, these can be estimated through energy
conservation arguments.

\section*{Acknowledgements}

We thank Alexander Brown, Evert Glebbeek, and Ujwal Kharel for useful
discussions. We also acknowledge the use of the 3D virtual
collaboration environment {\tt Qwaq Forums}, which greatly sped up the
remote collaboration leading to this paper. The calculation for this
work have been carried out on LISA cluster, which is hosted by SARA
supercomputing centre. JCL is supported by the National Science
Foundation under Grant No.\ 0703545, and EG and SPZ are supported by
NWO (grants \#635.000.303 and \#643.200.503).  This project has been
started during the Modest-7f workshop, which was hosted by the
University of Amsterdam and supported by NOVA, NWO, LKBF and ASTROSIM
grants.

\bibliographystyle{mn2e} \bibliography{GLPZ2007l}

\begin{thebibliography}{}

\bibitem[\protect\citeauthoryear{{Clayton}}{{Clayton}}{1983}]{1983psen.book...%
..C}
{Clayton} D.~D.,  1983, {Principles of stellar evolution and nucleosynthesis}.
Chicago: University of Chicago Press, 1983

\bibitem[\protect\citeauthoryear{{Dale} \& {Davies}}{{Dale} \&
  {Davies}}{2006}]{2006MNRAS.366.1424D}
{Dale} J.~E.,  {Davies} M.~B.,  2006, \mnras, 366, 1424

\bibitem[\protect\citeauthoryear{Dimotakis}{Dimotakis}{2005}]{2005ARFM...37..3%
29D}
Dimotakis P.~E.,  2005, Annual Review of Fluid Mechanics, 37, 329

\bibitem[\protect\citeauthoryear{{Freitag}, {G{\"u}rkan} \& {Rasio}}{{Freitag}
  et~al.}{2006}]{2006MNRAS.368..141F}
{Freitag} M.,  {G{\"u}rkan} M.~A.,    {Rasio} F.~A.,  2006, \mnras, 368, 141

\bibitem[\protect\citeauthoryear{{Gaburov}, {Gualandris} \& {Portegies
  Zwart}}{{Gaburov} et~al.}{2007}]{2007arXiv0707.0406G}
{Gaburov} E.,  {Gualandris} A.,    {Portegies Zwart} S.,  2007, ArXiv e-prints,
  707

\bibitem[\protect\citeauthoryear{{Lombardi}, {Thrall}, {Deneva}, {Fleming} \&
  {Grabowski}}{{Lombardi} et~al.}{2003}]{2003MNRAS.345..762L}
{Lombardi} J.~C.,  {Thrall} A.~P.,  {Deneva} J.~S.,  {Fleming} S.~W.,
  {Grabowski} P.~E.,  2003, \mnras, 345, 762

\bibitem[\protect\citeauthoryear{{Lombardi} Jr., {Proulx}, {Dooley},
  {Theriault}, {Ivanova} \& {Rasio}}{{Lombardi}
  et~al.}{2006}]{2006ApJ...640..441L}
{Lombardi} Jr. J.~C.,  {Proulx} Z.~F.,  {Dooley} K.~L.,  {Theriault} E.~M.,
  {Ivanova} N.,    {Rasio} F.~A.,  2006, \apj, 640, 441

\bibitem[\protect\citeauthoryear{{Lombardi} Jr., {Rasio} \&
  {Shapiro}}{{Lombardi} et~al.}{1996}]{1996ApJ...468..797L}
{Lombardi} Jr. J.~C.,  {Rasio} F.~A.,    {Shapiro} S.~L.,  1996, \apj, 468, 797

\bibitem[\protect\citeauthoryear{{Lombardi} Jr., {Warren}, {Rasio}, {Sills} \&
  {Warren}}{{Lombardi} et~al.}{2002}]{2002ApJ...568..939L}
{Lombardi} Jr. J.~C.,  {Warren} J.~S.,  {Rasio} F.~A.,  {Sills} A.,    {Warren}
  A.~R.,  2002, \apj, 568, 939

\bibitem[\protect\citeauthoryear{Monaghan}{Monaghan}{2005}]{0034-4885-68-8-R01}
Monaghan J.~J.,  2005, Reports on Progress in Physics, 68, 1703

\bibitem[\protect\citeauthoryear{{Paxton}}{{Paxton}}{2004}]{2004PASP..116..699%
P}
{Paxton} B.,  2004, \pasp, 116, 699

\bibitem[\protect\citeauthoryear{{Portegies Zwart}, {Baumgardt}, {Hut},
  {Makino} \& {McMillan}}{{Portegies Zwart} et~al.}{2004}]{2004Natur.428..724P}
{Portegies Zwart} S.~F.,  {Baumgardt} H.,  {Hut} P.,  {Makino} J.,
  {McMillan} S.~L.~W.,  2004, \nat, 428, 724

\bibitem[\protect\citeauthoryear{{Portegies Zwart}, {Makino}, {McMillan} \&
  {Hut}}{{Portegies Zwart} et~al.}{1999}]{1999A&A...348..117P}
{Portegies Zwart} S.~F.,  {Makino} J.,  {McMillan} S.~L.~W.,    {Hut} P.,
  1999, \aap, 348, 117

\bibitem[\protect\citeauthoryear{{Sills}, {Adams}, {Davies} \& {Bate}}{{Sills}
  et~al.}{2002}]{2002MNRAS.332...49S}
{Sills} A.,  {Adams} T.,  {Davies} M.~B.,    {Bate} M.~R.,  2002, \mnras, 332,
  49

\bibitem[\protect\citeauthoryear{{Springel}}{{Springel}}{2005}]{2005MNRAS.364.%
1105S}
{Springel} V.,  2005, \mnras, 364, 1105

\bibitem[\protect\citeauthoryear{{Suzuki}, {Nakasato}, {Baumgardt},
  {Ibukiyama}, {Makino} \& {Ebisuzaki}}{{Suzuki}
  et~al.}{2007}]{2007astro.ph..3290S}
{Suzuki} T.~K.,  {Nakasato} N.,  {Baumgardt} H.,  {Ibukiyama} A.,  {Makino} J.,
     {Ebisuzaki} T.,  2007, ArXiv Astrophysics e-prints

\bibitem[\protect\citeauthoryear{{Turner}, {Chapman}, {Bhattal}, {Disney},
  {Pongracic} \& {Whitworth}}{{Turner} et~al.}{1995}]{1995MNRAS.277..705T}
{Turner} J.~A.,  {Chapman} S.~J.,  {Bhattal} A.~S.,  {Disney} M.~J.,
  {Pongracic} H.,    {Whitworth} A.~P.,  1995, \mnras, 277, 705

\end{thebibliography}

\end{document}